# Availability Assessment of SunOS/Solaris Unix Systems based on Syslogd and wtmpx log files: A case study

Cristina Simache and Mohamed Kaâniche
*LAAS-CNRS — 7 Avenue du Colonel Roche*
*31077 Toulouse Cedex 4 — France*
*Mohamed.Kaaniche@laas.fr*

## Abstract

*This paper presents a measurement-based availability assessment study using field data collected during a 4-year period from 373 SunOS/Solaris Unix workstations and servers interconnected through a local area network. We focus on the estimation of machine uptimes, downtimes and availability based on the identification of failures that caused total service loss. Data corresponds to* `syslogd` *event logs that contain a large amount of information about the normal activity of the studied systems as well as their behavior in the presence of failures. It is widely recognized that the information contained in such event logs might be incomplete or imperfect. The solution investigated in this paper to address this problem is based on the use of auxiliary sources of data obtained from* `wtmpx` *files maintained by the SunOS/Solaris Unix operating system. The results obtained suggest that the combined use of* `wtmpx` *and* `syslogd` *log files provides more complete information on the state of the target systems that is useful to provide availability estimations that better reflect reality.*

## 1. Introduction

Event logs have been widely used to analyze the error/failure behavior of computer-based systems and to estimate their dependability. Event logs include a large amount of information about the occurrence of various types of events that are collected concurrently with normal system operation, and as such reflect actual workload and usage. Some of the events are informational and are issued from the normal activity of the target systems, whereas others are recorded when errors and failures affect local or distributed resources, or are related to system shutdown and start-up. The latter events are particularly useful for dependability analysis.

Computer system dependability analysis based on event logs has been the focus of several published papers [1, 2, 4, 5, 7, 8, 9]. Various types of systems have been studied (Tandem, VAX/VMS, Unix, Windows NT, Windows 2000, etc.) including mainframes and largely deployed commercial systems. The issues addressed in these studies cover a large spectrum, including the development of techniques and methodologies for the extraction of relevant information from the event logs, the identification of error patterns, their causes and their effects, and the statistical assessment of dependability measures such as failure and recovery rates, reliability and availability. It is widely recognized that such event log based dependability analyses provide useful feedback to software and system designers. Nevertheless, it is important to note that the results obtained are intimately related to the quality and the accuracy of the data recorded in the logs. The study reported in [1] points out various problems that might affect the data included in the event logs and make incorrect conclusions likely, considering as an example the VAX/VMS system. Thus, extreme care is needed to identify deficiencies in the data and to avoid that they lead to incorrect conclusions.

In this paper, we show that similar problems can be observed in the event logs maintained by the SunOS/Solaris Unix operating system, and we present a novel approach that is aimed to address such problems and to improve the dependability estimates based on such event logs. These results are illustrated using field data collected during a 4-year period from 373 SunOS/Solaris Unix workstations and servers interconnected through a LAN. The data corresponds to event logs recorded via the `syslog` daemon. In particular, we use `var/adm/messages` log files. We focus on the evaluation of machine uptimes, downtimes and availability based on the identification of failures that caused a total service interruption of the machine. In this study, we show that the consideration of the information recorded in the `var/adm/messages` log files only may lead to dependability estimations that do not faithfully reflect reality due to incomplete or imperfect data recorded in the corresponding logs. For the estimation of these measures, we start with the assumption that machine failures can be identified by the last events recorded in the event log before the machine goes down and then is rebooted. This assumption was considered in the study reported in [3]. However, the validity of this assumption is questionable in the following situations: 1) the machine has a real activity between the last event logged and the reboot without generating events in the logs, 2) the time when the failure occurs is earlier than the timestamp of the last event logged on the machine. To address these problems and to obtain more realistic estimations, we propose a solution based on utilization of additional information obtained from `wtmpx` Unix files, as well as data characterizing the state of the machines included in the data collection that are recorded at a regular basis during the data collection procedure. The results clearly show that the combined use of this additional information and `syslogd` log files have a significant impact on the estimations.

To our knowledge, the approach discussed in this paper and the corresponding results have not been addressed in the previous studies published on the exploitation of `syslogd` log files for the dependability analysis of Unix based systems, including our paper [4].

The rest of the paper is structured into 5 sections. Section 2 describes the event logging mechanism in Unix and the data collection procedure that we have used in our study. Section 3 presents the dependability measures that we have considered and discusses different approaches and assumptions to estimate them from the collected data. Section 4 presents some results illustrating the benefits of the proposed approach, as well as various statistics characterizing the dependability of the Unix systems considered in our study.

## 2. Event logging and data collection

For the Unix operating system, the event logging mechanism is implemented by the `syslog` daemon (denoted as `syslogd`). Running as a background process, this daemon listens for the events generated by different sources: kernel, system components (disk, memory, network interfaces), daemons and applications that are configured to communicate with `syslogd`. These events inform about the normal activity of the system as well as its behavior under the occurrence of errors and failures including reboot and shutdown events. The configuration file `/etc/syslog.conf` specifies the destination of each event received by `syslogd`, depending on its severity level and its origin. The destination could be one or several log files, the administration console or the operator.

The events that are relevant to our study are generally stored in the `/var/adm/messages` log file. Each message stored in a log file refers to an event that occurred on the system due to the local activity or its interaction with other systems on the network. It contains the following information: the date and time of the event, the machine name on which the event is logged and a description of the message. An example of an event recorded in the log file is given below:
`Mar 2 10:45:12 elgar automountd[124]: server mahler not responding`

The SunOS/Solaris Unix operating system limits the size of the log files. Generally, only the log files corresponding to the last 5 weeks of activity are kept. It is necessary to set up a data collection strategy in order to archive a large amount of data. This is essential to obtain representative results for the dependability measures characterizing the monitored systems.

In our study, we have included all the SunOS/Solaris machines connected through the LAAS local area network, excluding those used for experimental testbeds or maintenance activities. We have developed a data collection strategy to automatically collect the `/var/adm/messages` log files stored on these machines. This strategy takes into account the frequent evolution of the network configuration during the observation period in terms of variation of the number of connected systems, updates or changes of the operating system versions, modification of software configurations, etc. A shell script executed each week via the `cron` mechanism implements the strategy and remotely copies the log files from each system included in the study and archives them on a dedicated machine. After each data collection campaign, a text file (named `DCSummary`) containing a summary of the data collection campaign is created. This summary indicates the status of each machine included in the campaign and how the collection of the corresponding log file has been done.

For each machine, the status information reported in the summary is one of the following:
- *alive_OK*: the machine is alive and the copy of its log file succeeded;
- *alive_KO*: the machine is alive but the copy of its log file failed. For this case, a description of the failure symptom and cause is also included: shell problem, connection ended by tiers, etc.
- *no_answer*: the machine did not answer to a *ping* request before expiration of the default timeout period.

The information included in the `DCSummary` file is used to verify each data collection campaign and solve the problems that may appear during the collection. It is also useful to improve the accuracy of dependability measures estimation (see Section 3.2). More detailed information about the `syslogd` mechanism and the data collection strategy are reported in [6].

## 3. Dependability measures estimation and assumptions

Various types of dependability analyses can be carried out based on the information contained in the log files and several quantitative measures can be considered to characterize the dependability of the target machines: machine uptimes and downtimes, reliability, availability, failure and recovery rates, etc. In order to evaluate these measures, it is necessary to identify from the log files the failure occurrences and the corresponding service degradation durations. Such task is tedious and requires the development of heuristics and predefined failure criteria. An example of such analysis is reported in [7].

In our study, we have focused on the availability analysis of the individual machines included in the data collection. In this context, we have considered machine failures leading to a total interruption of the service delivered to the users, followed by a reboot. The time between the failure occurrence and the end of the reboot corresponds to the *total service interruption period* of the system. Apart from these periods, the system is considered to be in the normal functioning state where it delivers an appropriate service to the users.

In order to evaluate the availability of the machines included in the study, we need to estimate for each machine the corresponding uptimes (denoted as $UTi$) and downtimes ($DTi$), based on the information recorded in the event logs. Each downtime value $DTi$ corresponds to the *total service interruption period* associated to the $i^{th}$ failure. It is composed by the service degradation period due to the failure occurrence and the reboot period. Each uptime value corresponds to the period between two successive downtimes.

Using the uptime and downtime estimates for each machine $j$, we can evaluate the corresponding availability (noted $A_j$) and the unavailability (noted $UA_j$). These measures are computed with the following formulas:

$$UA_j = \Sigma UTi / \Sigma (UTi + DTi) \quad \text{and} \quad UA_j = 1 - UA_j \quad (1)$$

### 3.1. Machine uptimes and downtimes estimation

The estimation of machine uptimes and downtimes is carried out in two steps:
1) Identification of machine reboots and their duration.
2) Identification of failures associated to each reboot and of the corresponding service interruption period.

To identify the occurrence of machine reboots and their duration, we have developed an algorithm based on the sequential parsing and matching of each event recorded in the system log files to specific patterns or sequences of patterns characterizing the occurrence of reboots. Indeed, whereas some reboots can be explicitly identified by a "reboot" or a "shutdown" event, many others can be detected only by identifying the sequence of the initialization events that are generated by the system when it is restarted. The algorithm is described in [4, 6]. It gives, for each reboot $i$ identified in the event logs and for each machine, the timestamp of the reboot start ($dateSBi$), the timestamp of the reboot end ($dateEBi$) and the associated service interruption duration.

The identification of the timestamp of the failure associated to each reboot and the corresponding service interruption period is more problematic. In the study reported in [3], it was assumed that the timestamp of the last event recorded before the reboot (denoted as $dateEBR_i$) identifies the failure occurrence time. With this assumption, each uptime $UTi$ and downtime $DTi$ can be evaluated as follows:

$$UTi = dateEBR_i - dateEB_{i-1} \text{ and}$$
$$DTi = dateEB_i - dateEBR_i \quad (2)$$

where $i$ is the index of the current reboot, $i-1$ the index of the previous reboot.

The consideration of EBR for the estimation of $UTi$ and $DTi$ parameters may not be realistic in the following situations (denoted as S1 and S2):
S1) The system could be in a normal functioning state during a period of time between EBR and the following reboot although it does not generate any event into the log files during that period.
S2) The beginning of the service interruption period for the users could be prior to the timestamp of the EBR event. This happens for instance when a critical failure affects the machine in such a way that it becomes completely unusable to the users, without preventing the event logging mechanisms from recording some messages into the log files.

A careful analysis of the data collected during our study revealed that the above situations are common. To address this problem and to improve downtime and uptime estimation accuracy, it is necessary to use auxiliary data that provides complementary information on the activity of the target machines.

In this paper, we present a solution based on the correlation of data collected from the `/var/adm/messages` log files, with data issued from `wtmpx` files also maintained by the SunOS/Solaris operating system. We also use the information recorded in the `DCSummary` file (see Section 2). The following section presents the method developed to extract the data from the `wtmpx` file and how we used this data to adjust the estimation of machine uptimes and downtimes.

## 3.2. Uptime and downtime estimations refinement

**3.2.1. `wtmpx` files.** The SunOS/Solaris Unix operating system records into the `/var/adm/wtmpx` binary file information identifying the users login/logout. Through the pseudo-user *reboot* it also records information on the system reboots. The `wtmpx` file is organized into records (named also entries) with a fixed size. Each record has the format of a data structure with the following fields:
- the user login name: "*user*";
- the id associated to the current record in the /etc/inittab file: "*init_id*";
- the device name (console, lnxx): "*device*";
- the process id: "*pid*";
- the record type: "*proc_type*";
- the exit status for a process marked as DEAD_PROCESS: "*exit_status*" and "*term_status*";
- the timestamp of the record: "*date*";
- the session id: "*session_id*";
- the length of the machine's name: "*length*";
- the machine's name used by the user to connect, if it is a remote one: "*host*".

We developed a specific algorithm that collects the `wtmpx` file of each machine included in the study on a regular basis and processes the binary file to extract the information that is relevant to our study. The results of the algorithm are kept in a separate file for each machine. Figure 1 presents examples of records obtained for a machine of our network.

The first two records show that the *root* user connected to the local system from the system named *cubitus* on November 6, 2001 at 16h 37mn 41s, using the *rlogin* command. The next records inform about the occurrence of a reboot event about 3 minutes later. The third record shows that this reboot was done via a *shutdown* command executed probably by the root user. The sequence of records corresponding to a reboot event is much longer than this example. The whole sequence is not presented in Figure 1, the aim of the illustration is to show some examples of records as extracted from `wtmpx` files by our algorithm.

In the following, we outline the approach that we developed to use the information extracted from the `wtmpx` files together with the information from the `DCSummary` files in order to refine the uptime and downtime estimations, considering situations S1 and S2 discussed in Section 3.1.

| | | |
|---|---|---|
| 2001 Nov 6 16:37:41 | user=.rlogin<br>length=0<br>device=/dev/pts/1<br>proc_type=6 | host=<br>init_id=r100<br>pid=25220<br>term_status=0 |
| 2001 Nov 6 16:37:41 | user=root<br>length=8<br>device=/dev/pts/1<br>proc_type=7 | host=cubitus<br>init_id=r100<br>pid=25220<br>term_status=0 |
| 2001 Nov 6 16:40:35 | user=shutdown<br>length=0<br>device=~<br>pid=0<br>term_status=0 | host=<br>init_id=<br><br>proc_type=0<br>exit_status=0 |
| 2001 Nov 6 16:41:39 | user=<br>length=0<br>device=system boot<br>proc_type=2 | host=<br>init_id=<br>pid=0<br>term_status=0 |
| 2001 Nov 6 16:42:09 | user=<br>length=0<br>device=run–level 3<br>proc_type=1 | host=<br>init_id=<br>pid=0<br>term_status=0 |

**Figure 1. Examples of records from /var/adm/wtmpx obtained with our algorithm**

**3.2.2. Situation S1: an operational activity exists between EBR and SB events.** The detailed analysis of the collected data from the log files and comparison with the information extracted from `wtmpx` files showed that the situation where a real activity exists between the last event recorded before a reboot (EBR) and the event identifying the start of the following reboot (SB event) recorded in the `/var/adm/messages` log files appears quite often. This situation occurs when the machine functions normally but its activity doesn't produce any message into the log file maintained by the syslogd daemon. The cause could be that the applications or services run by the users aren't configured to communicate with the syslogd daemon.

To better understand this case, Figure 2 gives an example of a sequence of events characterizing the state of the corresponding system, taking into account the information extracted from the `/var/adm/messages`, `wtmpx` and `DCSummary` files. For each event, we indicate the timestamp when it is logged, a short description and the source file from which the event is extracted. For `wtmpx` events, we present only the fields which are useful to identify the system activity, the other fields are not significant for this analysis.

For this example, the events recorded in the `/var/adm/messages` log file let us believe that the system had no activity between December 8 at 18:06 (EBR event) and December 9 at 15:30, the timestamp of the reboot start. However, the analysis of the `DCSummary` and `wtmpx` files shows that the system had a real activity between EBR and SB events. In fact, we see that the data collection campaign was successfully carried out on December 9 at 6:43.

| Event # | Event date | Event description | | | | File where the event is logged |
|---|---|---|---|---|---|---|
| 1 | 2002 Dec 8 18:06:08 | last event before reboot <EBR> | | | | var/adm/messages log file |
| 2 | 2002 Dec 9 06:43:34 | alive_ok | | | | DCSummary |
| 3 | 2002 Dec 9 13:18:45 | *user*=UserC; | *device*=pts/0; | *pid*=2362; | *proc_type*=7 | wtmpx |
| 4 | 2002 Dec 9 13:35:21 | *user*=UserB; | *device*=pts/1; | *pid*=2379; | *proc_type*=7 | wtmpx |
| 5 | 2002 Dec 9 13:47:57 | *user*=UserB; | *device*=pts/1; | *pid*=2379; | *proc_type*=8 | wtmpx |
| 6 | 2002 Dec 9 13:48:48 | *user*=UserA; | *device*=pts/1; | *pid*=2434; | *proc_type*=7 | wtmpx |
| 7 | 2002 Dec 9 15:18:46 | *user*=UserA; | *device*=pts/1; | *pid*=2434; | *proc_type*=8 | wtmpx |
| 8 | 2002 Dec 9 15:29:20 | *user*=UserB; | *device*=console; | *pid*=2644; | *proc_type*=7 | wtmpx |
| 9 | 2002 Dec 9 15:29:25 | *user*=UserB; | *device*=console; | *pid*=338; | *proc_type*=8 | wtmpx |
| 10 | 2002 Dec 9 15:29:25 | *user*=UserB; | *device*=console; | *pid*=2644; | *proc_type*=8 | wtmpx |
| 11 | 2002 Dec 9 15:29:27 | *user*=LOGIN; | *device*=console; | *pid*=2742; | *proc_type*=6 | wtmpx |
| ......... | ......... | ......... | | | | ......... |
| 12 | 2002 Dec 9 15:29:52 | *user*=troot; | *device*=console; | *pid*=334; | *proc_type*=7 | wtmpx |
| 13 | 2002 Dec 9 15:30:52 | *user*=sac; | *device*=; | *pid*=333; | *proc_type*=8 | wtmpx |
| 14 | 2002 Dec 9 15:30:52 | *user*=troot; | *device*=console; | *pid*=334; | *proc_type*=8 | wtmpx |
| 15 | 2002 Dec 9 15:30:52 | *user*=; | *device*=run-level 6; | *pid*=0; | *proc_type*=1 | wtmpx |
| 16 | 2002 Dec 9 15:30:52 | *user*=rc6; | *device*=; | *pid*=2899; | *proc_type*=5 | wtmpx |
| 17 | 2002 Dec 9 15:30:53 | reboot start <SB> | | | | var/adm/messages log file |

**Figure 2. Example illustrating situation S1**

Moreover, the records from wtmpx file show, for example, that *UserA* used the system on December 9 between 13:48 (information given by the *proc_type* field value equal to 7, that is the process with *pid*=2434 started at the time of this record) and 15:18 (proc_type=8, the same process ended at the time of this record), corresponding to an utilization period of the system of nearly one hour and a half.

In this situation, the EBR event as defined earlier doesn't correspond to the beginning of the total service interruption period. Thus, the estimated value of the downtime parameter using the assumption discussed in Section 3.1, does not faithfully reflect the real value of the service interruption period. Based on the correlation of the information provided by the three data source files, a refined and more accurate estimation of machine downtimes and uptimes could be obtained. The refinement consists in associating the failure occurrence time to the timestamps of the last event recorded before the reboot based on the information contained in /var/adm/messages, wtmpx and DCSummary files.

**3.2.3. Situation S2: the service interruption period starts before the EBR event.** This situation occurs when critical failures affect the system in such a way that it becomes completely unusable, without preventing the event logging mechanisms from recording some messages into the log files. During the recovery phase, the actions performed by the system administrators may include several unsuccessful reboot attempts that are not recorded in the /var/adm/messages log file, but some events referring to them are written in the wtmpx file. Using this information, just like in the previous case, we can refine the downtime and uptime estimations by associating the failure occurrence time to the timestamps of the events recorded in the wtmpx file that better reflects the start of the service interruption. An example of a sequence of events illustrating this case is given in Figure 3.

| Event # | Event date | Event description | | | | File where the event is logged |
|---|---|---|---|---|---|---|
| 1 | 2003 Jan 9 10:18:59 | *user*=root; | *device*=console; | *pid*=2370; | *proc_type*=7 | wtmpx |
| 2 | 2003 Jan 9 10:21:39 | *user*=sac; | *device*=; | *pid*=425; | *proc_type*=8 | wtmpx |
| 3 | 2003 Jan 9 10:21:39 | *user*=root; | *device*=console; | *pid*=2370; | *proc_type*=8 | wtmpx |
| 4 | 2003 Jan 9 10:21:39 | *user*=; | *device*=run-level 5; | *pid*=0; | *proc_type*=1 | wtmpx |
| 5 | 2003 Jan 9 10:21:39 | *user*=rc5; | *device*=; | *pid*=25952; | *proc_type*=5 | wtmpx |
| 6 | 2003 Jan 9 10:21:48 | *user*=UserC; | *device*=pts/3; | *pid*=11584; | *proc_type*=8 | wtmpx |
| 7 | 2003 Jan 9 10:21:48 | *user*=UserC; | *device*=pts/1; | *pid*=11359; | *proc_type*=8 | wtmpx |
| 8 | 2003 Jan 9 10:22:05 | last event before reboot <EBR> | | | | var/adm/messages log file |
| 9 | 2003 Jan 9 10:22:13 | *user*=rc5; | *device*=; | *pid*=25953; | *proc_type*=8 | wtmpx |
| 10 | 2003 Jan 9 10:22:13 | *user*=uadmin; | *device*=; | *pid*=26121; | *proc_type*=5 | wtmpx |
| 11 | 2003 Jan 9 10:22:16 | reboot start <SB> | | | | var/adm/messages log file |

**Figure 3. Example illustrating situation S2**

We can identify the events extracted from the `wtmpx` file informing upon the stop of the system: event # 2 with *user* field "sac" and *proc_type* " 8" (dead process) followed by events #3, #4, and #5 notifying the system run-level change to run-level 5 (this one is used to properly stop the system).

This example shows that the start of the service interruption period is prior to the EBR event recorded in the `/var/adm/messages` log file. The refinement of the uptime and downtime estimations corresponding to such situations consists in associating the failure occurrence time to the timestamps of the last event recorded in the `wtmpx` file before the start of the reboot sequence.

## 4. Experimental results

The analyses presented in this Section are based on `/var/adm/messages` log file data collected during 45 months (October 1999 – July 2003) from 418 SunOS/Solaris Unix workstations and servers interconnected through the LAAS local area computing network. As shown in Figure 4, the data collection period differed significantly from one machine to another due to the dynamic evolution of the network. For more than 70 % of the machines, the data collection period was longer than 21 months. On the other hand, it can be noticed that some machines have a quite short data collection period. In order to have significant statistical analysis results, we excluded from the analysis the machines for which the data collection period was shorter than 2000 hours (about 3 months). Consequently, the results presented in the following concern 373 Unix machines. Among these machines, 17 correspond to major servers for the entire network or a sub-set of users: WWW, NIS+, NFS, FTP, SMTP, file servers, printing servers, etc.

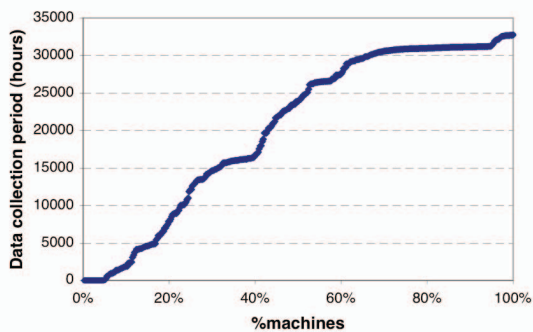

**Figure 4. Examples of records from /var/adm/wtmpx obtained with our algorithm**

The application of the reboot identification algorithm on the collected data allowed us to identify 12805 reboots for the 373 machines, only 476 reboots concern the 17 servers. Based on the information provided by the reboot identification algorithm, we evaluated for each machine the associated uptimes $UT_i$ and downtimes $DT_i$, and the availability measure.

The collection of `wtmpx` files started later than the `/var/adm/messages` log files. For this reason, we were able to analyze the impact of uptimes and downtimes estimation refinement algorithms only on a subset of $UT_i$ and $DT_i$ values associated with the reboots identified from the log files. Among the 12805 reboots, this analysis concerned 6163 reboots (48.13%). For the remaining 6642 reboots, the corresponding data from the `wtmpx` files was not available.

In the following, we first present in Section 4.1 the results of machine uptimes and downtimes estimation based on the processing of the set of 6163 reboots focusing on the impact of the estimation refinement algorithms. Then, global results taking into account the whole data collected during our study are presented in Section 4.2 in order to give an overall picture on the availability and the rate of occurrence of reboots characterizing the Unix machines included in our study.

### 4.1. Machine uptimes and downtimes estimation and refinement

The correlation of the information contained in the `/var/adm/messages` log files, the `wtmpx` files, and the `DCSummary` files, revealed that both situations S1 and S2 discussed in Section 3.2 are common:
- Situation S1 was observed for 79.35% of the analyzed reboots;
- Situation S2 was observed for 10.77% of the analyzed reboots;

For the 9.88% remaining reboots, the assumption that the EBR recorded in `/var/adm/messages` file identifies the last event recorded on the machine before the reboot was consistent with the information available in the `wtmpx` and the `DCSummary` files.

In order to analyze the impact of the estimation refinement algorithms on the results, Table 1 gives the Mean, Median and Standard Deviation of uptime and downtime values, before and after the application of our estimation refinement algorithms discussed in Section 3. Considering the median of the downtime values, it can be seen that the refinement algorithms have a significant impact on the results. The median estimated after the refinement is 66 times lower than the value obtained without the refinement. The refinement algorithms have also an impact on the uptimes estimation, but as expected the improvement factor is lower than the one observed for the downtime values (1.8 compared to 66).

**Table 1. Machine uptimes and downtimes estimates before and after refinement**

|  | Uptimes UTi | | Downtimes DTi | |
|---|---|---|---|---|
|  | before refinement | after refinement | before refinement | after refinement |
| Mean | 28.3 days | 1.1 month | 5.9 days | 1.9 days |
| Median | 6.1 days | 10.8 days | 8.9 hours | 8.1 min |
| Std. Dev. | 1.7 months | 1.8 months | 24.1 days | 21.1 days |

The impact of the estimation refinement algorithms on availability is summarized in Table 2. It can be seen the estimated average unavailability after the refinement is three times lower than the value estimated based only on the information in the `/var/adm/messages` log files. Clearly, the difference is significant and cannot be ignored.

**Table 2. Impact of the estimation refinement algorithms on Availability and Unavailability**

|  | before refinement | after refinement |
|---|---|---|
| A | 89.3% | 96.3 % |
| UA | 39.0 days/year | 13.7 days/year |

### 4.2. Availability and reboot rates estimated from the whole data set

This section presents some results concerning the reboot rates and the availability of the 373 SunOS/Solaris Unix machines included in our study taking into account the whole set of 12805 reboots identified from the `/var/adm/messages` files. When the `wtmpx files` were not available (this concerned 6642 reboots), the estimation of the $UTi$, $DTi$, availability and reboot rates was based only on the information in the `/var/adm/messages` files, using the assumption discussed in Section 3.1. In the other case (i.e., for the 6163 reboots), we applied the estimation refinement algorithms presented in Section 3.2.

Figure 5 plots the reboot rates estimated for each machine as a function of the data collection period. The estimated reboot rate for each machine corresponds to the average number of reboots recorded during the corresponding observation. It can be seen that the reboot rates are uniformly distributed between $10^{-4}$/hour and $10^{-2}$/hour.

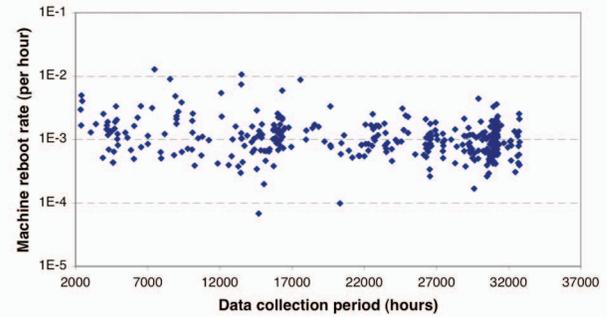

**Figure 5. Unix machines reboot rates as a function of the data collection period**

As indicated in Table 3, the mean value of machine reboot rates is $1.3\ 10^{-3}$/hour, when considering all Unix machines including workstations and servers. If we take into account only the servers, the mean reboot rate is 1.5 times lower ($7.7\ 10^{-4}$/hour) corresponding to one reboot every two months.

**Table 3. Reboot rate statistics**

|  | Mean | Median | Std. Dev. |
|---|---|---|---|
| SunOS/Solaris machines (Workstations + Servers) | $1.3\ 10^{-3}$/h | $1.0\ 10^{-3}$/h | $1.3\ 10^{-3}$/h |
| Servers only | $7.7\ 10^{-4}$/h | $6.4\ 10^{-4}$/h | $5.6\ 10^{-4}$/h |

The results illustrating the availability and unavailability of the Unix machines including workstations and servers are given in Figure 6 and Table 4. The mean availability is 97.81 % corresponding to an average unavailability of 8 days per year. Detailed analysis shows that only 15 among the 373 Unix machines included in the study have an availability lower than 90%.

When considering only the servers, the estimated availability varies between 99.36% and 99.1% with an average unavailability of 12 hours per year.

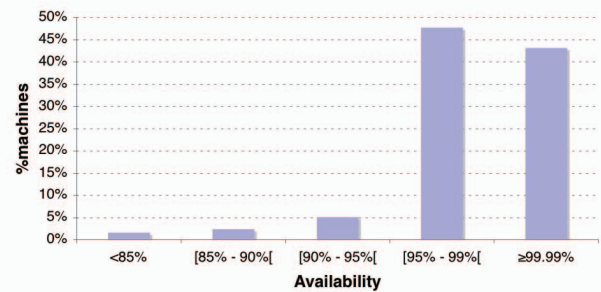

**Figure 6. SunOS/Solaris Unix machines availability distribution**

**Table 4. Availability and Unavailability statistics**

|    | Mean         | Median       | Std. Dev.     |
|----|--------------|--------------|---------------|
| A  | 97.81 %      | 98.79 %      | 3.07 %        |
| UA | 7.99 day/year| 4.41 day/year| 11.20 day/year|

## 6. Conclusion

Dependability analyses based on event logs provide useful feedback to software and system designers. Nevertheless, the results obtained are intimately related to the quality and the completeness of the information recorded in the logs. As the information contained in such event logs could be incomplete or imperfect, it is important to use additional sources of information to ensure that the conclusions derived from such analyses faithfully reflect reality. The approach investigated in this paper is aimed to fulfill this objective considering SunOS/Solaris Unix systems as an example.

In particular, we have shown that the combined us of the data contained in the `syslogd` files and the information recorded in the `wtmpx` files or through the monitoring of systems state during the data collection campaigns provides uptime and downtime estimations that are closer to reality than the estimations obtained based on `syslogd` files only. This result is illustrated based on a large set of field data collected from 373 machines during a 45 month observation period.

In our future work, we will investigate the applicability of the approach proposed in this paper to other operating systems such as Linux, Windows 2K and Mac OS X.